\documentclass[english,twocolumn,aps,showpacs,superscriptaddress,amsmath,amssymb,preprintnumbers]{revtex4}
\usepackage[]{fontenc}
\usepackage[latin1]{inputenc}
\usepackage{amsmath}
\usepackage{graphicx}
\usepackage{amssymb}

\makeatletter

\makeatletter
\usepackage{epsfig}

\makeatother

\usepackage{babel}
\makeatother
\begin{document}

\title{Pattern Selection in the Complex Ginzburg-Landau Equation with Multi-Resonant
Forcing }

\author{Jessica Conway and Hermann Riecke}

\affiliation{Engineering Sciences and Applied Mathematics, Northwestern University,
Evanston, IL 60208, USA}


\begin{abstract}
We study the excitation of spatial patterns by resonant, multi-frequency
forcing in systems undergoing a Hopf bifurcation to spatially homogeneous
oscillations. Using weakly nonlinear analysis we show that for small
amplitudes only stripe or hexagon patterns are linearly stable, whereas
square patterns and patterns involving more than three modes are unstable.
In the case of hexagon patterns up- and down-hexagons can be simultaneously
stable. The third-order, weakly nonlinear analysis predicts stable
square patterns and super-hexagons for larger amplitudes. Direct simulations
show, however, that in this regime the third-order weakly nonlinear
analysis is insufficient, and these patterns are, in fact unstable.
\end{abstract}
\maketitle
The variety of patterns and planforms that have been observed in surface
waves on vertically vibrated fluid surfaces (Faraday waves) is remarkable
\cite{ChAl92,EdFa93,KuPi98,ArFi02}. As elucidated in various theoretical
investigations \cite{ZhVi96,SiTo00,RuSi07} this extreme variability
is a result of the fact that the wave form of the vibration's forcing
function allows for a detailed tuning of various aspects of the interaction
between different plane-wave modes, which can stabilize complex patterns
like super-lattice patterns and quasi-patterns. Motivated by this
richness of patterns we are investigating here the effect of time-periodic
forcing with different wave forms on systems undergoing a Hopf bifurcation
to spatially homogeneous oscillations.

In order to describe a forced Hopf bifurcation within a weakly nonlinear
framework the forcing must be sufficiently weak. For it to have a
qualitative effect on the system it must then include frequencies
that are close to one or more of the low-order resonances of the system,
i.e. its spectrum has to contain frequencies close to the Hopf frequency
$\omega_{h}$ itself (1:1-forcing), close to $2\omega_{h}$ (2:1-forcing),
or close to $3\omega_{h}$ (3:1-forcing). These strong resonances
lead to additional terms in the weakly nonlinear description and qualitatively
affect the system \cite{CoEm92a}. To wit, in the weakly nonlinear
regime the complex oscillation amplitude $A$ satisfies a complex
Ginzburg-Landau equation of the form

\begin{eqnarray}
\frac{dA}{dt} & = & (\mu+i\nu)A+(1+i\beta)\Delta A-(1+i\alpha)A|A|^{2}+\nonumber \\
 &  & +\gamma\bar{A}+\eta\bar{A}^{2}+\zeta.\label{eq:cgle_123forcing}\end{eqnarray}
The forcing terms $\zeta$, $\gamma\bar{A}$, and $\eta\bar{A}^{2}$
express the effect of forcing the system at the frequencies $\omega$,
$2\omega$, and $3\omega$, respectively, with $\omega=\omega_{h}+\frac{\nu}{2}$.
The parameter $\mu$ expresses the distance from the Hopf bifurcation,
which is shifted by a $\mathcal{O}(|\eta|^{2})$ compared to the unforced
case. Here we will focus on the case $\zeta=0$. To include the forcing
near the $1:1$-resonance one can eliminate the inhomogeneous term
$\zeta$ by shifting $A$ by the fixed-point solution $A_{0}$ satisfying
$\zeta=-(\mu+i\nu)A_{0}+(1+i\alpha)A_{0}|A_{0}|^{2}-\gamma\bar{A_{0}}-\eta\bar{A}_{0}^{2}$
and use $A_{0}$ instead of $\zeta$ as an external parameter \cite{CoRiunpub}.

As is apparent from (\ref{eq:cgle_123forcing}), the forced Hopf bifurcation
is described by an equation that is very similar to a two-component
reaction-diffusion equation. The only and significant difference is
the term involving $\beta$ which characterizes the dispersion of
unforced traveling wave solutions, which would be absent in the reaction-diffusion
context. It plays, however, an essential role in exciting patterns
with a characteristic wavenumber \cite{CoFr94} and cannot be omitted
(cf. (\ref{eq:critical}) below). Pattern selection in a general two-component
reaction-diffusion system has been studied in detail by Judd and Silber
\cite{JuSi00}, who find that in principle not only stripe and hexagon
patterns can be stable in such systems, but also super-square and
super-hexagon patterns. They also find that despite the large number
of parameters characterizing these systems surprisingly only few,
very special combinations of the parameters enter the equations determining
the pattern selection.

\begin{figure}[!tph]
\includegraphics[width=7cm,keepaspectratio]{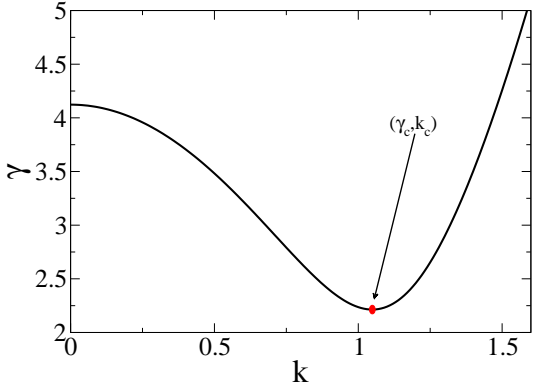}

\caption{Neutral stability curve for (\ref{eq:cgle_123forcing}) with $\mu=-1$,
$\beta=3$, $\nu=4$ and $\zeta=0$. The critical point $(k_{c,}\gamma_{c})=\left(\sqrt{\frac{11}{10}},\frac{7}{\sqrt{10}}\right)$
is marked by a red circle. \label{fig:NeutralCurve}}
\end{figure}

\textbf{Amplitude Equations.} In this paper we will stay below the
Hopf bifurcation taking $\mu<0$. Thus, as in Faraday systems, in
the absence of forcing, no oscillations arise. To investigate the
weakly nonlinear stable standing wave patterns possible in (\ref{eq:cgle_123forcing})
we derive amplitude equations for spatially periodic planforms. The
linear stability of the state $A=0$ is easily obtained by splitting
the equation and the amplitude $A$ into real and imaginary parts
($A\equiv A_{r}+iA_{i}$). The usual Fourier ansatz $A_{r,i}\propto e^{ikx}$
yields then the neutral stability curve $\gamma_{n}(k)$ with the
basic state being unstable for $\gamma>\gamma_{n}(k)$. The minimum
$\gamma_{c}(k)$ of the neutral curve is found to be at \begin{equation}
k_{c}^{2}=\frac{\mu+\nu\beta}{1+\beta^{2}},\:\qquad\gamma_{c}^{2}=\frac{(\nu-\mu\beta)^{2}}{(1+\beta^{2})}.\label{eq:critical}\end{equation}
Since $\mu<0$, the condition $k_{c}^{2}>0$ implies that spatial
patterns arise only if the detuning of the forcing relative to the
Hopf frequency is such that waves with non-zero $k$ are closer to
resonance than homogeneous oscillations with $k=0$ \cite{CoFr94}.
A typical neutral curve is illustrated in Figure \ref{fig:NeutralCurve}
for $\mu=-1$, $\nu=4$, $\beta=3$ and $\zeta=0.$ The weakly nonlinear
analysis presented in this paper is valid for values of $\gamma$
near $\gamma_{c}$. The range of validity is restricted by $\gamma_{n}(k=0)$
where spatially homogeneous oscillations are excited by the forcing,
which interact with the standing-wave modes with wavenumber $k_{c}$.

To determine the stability of the various planforms we first determine
the amplitude equations for rectangle patterns, which are comprised
of two modes separated by an angle $\theta$ in Fourier space. We
expand $(A_{r},A_{i})$ as\begin{equation}
\left(\begin{array}{c}
A_{r}\\
A_{i}\end{array}\right)=\epsilon\left\{ \sum_{j=1,\theta}Z_{j}(T)e^{i\mathbf{{\bf \textbf{k}}}_{j}\cdot{\bf \textbf{r}}}+c.c.\right\} \left(\begin{array}{c}
v_{1}\\
v_{2}\end{array}\right)+\mathcal{O}(\epsilon^{2}),\label{eq:RhWNLansatz}\end{equation}
 where $0<\epsilon\ll1$ and the complex amplitudes $Z_{1}(T)$ and
$Z_{\theta}(T)$ depend on the slow time $T=\epsilon^{2}t$. The wavevectors
are given by $\mathbf{{\bf \textbf{k}}_{1}}=(k_{c},0)$ and ${\bf \textbf{k}}_{\theta}=(k_{c}\cos(\theta),k_{c}\sin(\theta))$.
We also expand $\gamma$ as $\gamma=\gamma_{c}+\epsilon^{2}\gamma_{2}$.

The usual expansion leads to the amplitude equations for $(Z_{1},Z_{\theta})$,\begin{eqnarray}
\frac{dZ_{1}}{dT} & = & \lambda(\gamma-\gamma_{c})Z_{1}-\left(b_{0}|Z_{1}|^{2}+b_{1}(\theta)|Z_{\theta}|^{2}\right)Z_{1},\label{eq:RhombAmpEq1}\\
\frac{dZ_{\theta}}{dT} & = & \lambda(\gamma-\gamma_{c})Z_{\theta}-\left(b_{1}(\theta)|Z_{1}|^{2}+b_{0}|Z_{\theta}|^{2}\right)Z_{\theta}.\label{eq:RhombAmpEqstheta}\end{eqnarray}

If $\theta=\frac{n\pi}{3}$, $n\in\mathbb{Z}$, the quadratic nonlinearity
induces a secular term and the expansion has to include three modes
rotate by $120^{\circ}$ relative to each other,\begin{equation}
\left(\begin{array}{c}
A_{r}\\
A_{i}\end{array}\right)=\epsilon\left\{ \sum_{j=1}^{3}Z_{j}(T)\, e^{i{\bf \textbf{k}}_{j}\cdot\mathbf{r{\bf }}}+c.c.\right\} \left(\begin{array}{c}
v_{1}\\
v_{2}\end{array}\right)+\mathcal{O}(\epsilon^{2}).\label{eq:HexWNLansatz}\end{equation}
The parameters can be chosen such that the quadratic solvability condition
is delayed to cubic order, yielding\begin{eqnarray}
\frac{dZ_{1}}{dT} & = & \lambda(\gamma-\gamma_{c})Z_{1}+\sigma\bar{Z}_{2}\bar{Z}_{3}-\label{eq:HexAmpEqs}\\
 &  & -\left(b_{0}|Z_{1}|^{2}+b_{2}\left(|Z_{2}|^{2}+|Z_{3}|^{2}\right)\right)Z_{1},\nonumber \end{eqnarray}
with similar equations for $Z_{2}$ and $Z_{3}.$

More complex patterns can be described by combining these two analyses.
For example, a super-hexagon pattern comprised of two hexagon patterns
$\{ Z_{1},Z_{2},Z_{3}\}$ and $\{ Z_{4},Z_{5},Z_{6}\}$ that are rotated
relative to each other by an angle $\theta_{SH}$ is described by
the amplitude equation \begin{align*}
\frac{dZ_{1}}{dT}= & \lambda(\gamma-\gamma_{c})Z_{1}+\sigma\bar{Z}_{2}\bar{Z}_{3}-\\
 & -\left(b_{0}|Z_{1}|^{2}+b_{2}\left(|Z_{2}|^{2}+|Z_{3}|^{2}\right)\right)Z_{1}\\
 & -\sum_{j=0}^{2}b_{1}(\theta_{SH}+j\frac{2\pi}{3})|Z_{4+j}|^{2}Z_{1}\end{align*}
 and corresponding equations for $Z_{j}$, $j=2,\ldots,6$.

The coefficients of the amplitude equations can be written in a simple
form, setting $\eta\equiv\eta_{r}+i\eta_{i}$: \begin{equation}
\lambda=\frac{\sqrt{1+\beta^{2}}}{|\beta|},\;\;\sigma=\frac{2\sqrt{1+\beta^{2}}(a\eta_{r}+\eta_{i})}{\beta\sqrt{1+a^{2}}},\label{eq:lambda_sigma}\end{equation}
\begin{eqnarray}
b_{0} & = & 3\psi+\frac{76}{9}\chi,\label{eq:cubic}\\
b_{1}(\theta) & = & 6\psi+8f(\theta)\chi,\label{eq:b1}\\
b_{2} & = & 6\psi+10\chi+\phi\label{eq:b2}\end{eqnarray}
 with\begin{eqnarray}
\psi & = & \frac{-2a(\alpha-\beta)\sqrt{1+\beta^{2}}}{\beta(1+a^{2})},\label{eq:psi}\\
\chi & = & \frac{-\beta(\nu-\mu\beta)}{2(\mu+\nu\beta)}\sigma^{2},\label{eq:chi}\\
\phi & = & \frac{4(1+\beta^{2})^{\frac{3}{2}}\eta_{i}\left[\left(a+\beta\right)\eta_{r}-\left(a\beta-1\right)\eta_{i}\right]}{\beta^{2}(1+a^{2})(\nu-\mu\beta)}\label{eq:phi}\end{eqnarray}
and the angle dependence given by\[
f(\theta)=\frac{3+16\cos^{4}\theta}{(4\cos^{2}\theta-1)^{2}}.\]
 Here $a=\sqrt{1+\beta^{2}}+\beta$. The scaling of the nonlinear
coefficients is based on a normalized eigenvector ${\bf v}=\left(\begin{array}{c}
v_{1}\\
v_{2}\end{array}\right)$.

\textbf{Pattern Selection.} As shown by Judd and Silber \cite{JuSi00}
for general two-component reaction-diffusion systems, at the point
of degeneracy at which the quadratic coefficient $\sigma$ and with
it the coefficient $\chi$ vanishes not only stripe patterns but also
hexagon or triangle patterns can be stable. The conditions for hexagons
(or triangles) to be stable are\begin{eqnarray}
\phi & < & 0\label{eq:HexCond1}\\
-\frac{2\phi}{15}< & \psi & <-\frac{\phi}{3}.\label{eq:HexCond2}\end{eqnarray}
Whether hexagon or triangle patterns are stable depends on higher-order
terms in the amplitude equations \cite{SiPr98}, which are not considered
here. For $0<\psi<-\frac{2\phi}{15}$, hexagons are unstable to stripes.
Both patterns bifurcate unstably for $\psi<0$.

\begin{figure}[!tph]
\includegraphics[width=6cm]{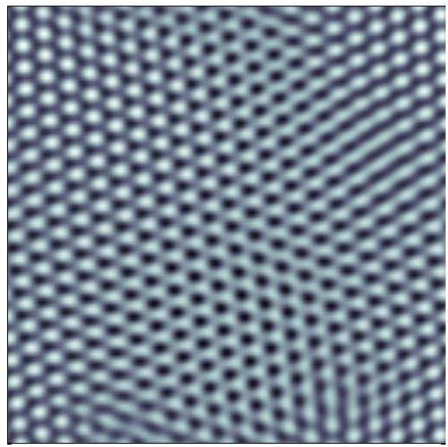}

\caption{Competing domains of up- and down-hexagon domains starting from random
initial conditions with linear parameters as in Fig. \ref{fig:NeutralCurve}
and nonlinear parameters $\alpha=-1$, $\eta_{r}=0.4$ and $\eta_{i}=0.4(\sqrt{10}+3)$.
$\eta_{r}$ and $\eta_{i}$ are chosen so that $\sigma=0.$}

\label{fig:up-and-down}
\end{figure}

Comparing the conditions (\ref{eq:HexCond1},\ref{eq:HexCond2}) with
expressions (\ref{eq:psi},\ref{eq:chi},\ref{eq:phi}) shows that
in the system considered here stripe patterns and hexagons can be
stable at the point of degeneracy, depending on the system parameters
$\alpha$ and $\beta$. Specifically, at the degeneracy one has $\eta_{i}=-a\eta_{r}$,
implying\begin{equation}
\phi(\sigma=0)=-\frac{4a\eta_{r}^{2}(1+\beta^{2})^{\frac{3}{2}}}{\beta(\nu-\mu\beta)}<0.\label{eq:phidegeneracy}\end{equation}
Here we have made use of the fact that $\beta(\nu-\mu\beta)>0$, which
follows from $\mu<0$ and the condition $k_{c}^{2}>0$. Whether condition
(\ref{eq:HexCond2}) is then satisfied depends on the specific unforced
system: only a certain range of values of $\alpha$ leads to stable
hexagons, which is determined by the $\alpha-$dependence of $\psi$.
A distinguishing feature of these hexagon patterns is that both `up'-
and `down'-hexagons are simultaneously stable and are likely to form
competing domains. Fig.\ref{fig:up-and-down} shows an example of
the competition between `up'- and `down'-hexagons in a numerical simulation
of (\ref{eq:cgle_123forcing}).

Unfolding the degeneracy, i.e. taking $0<|\sigma|\ll1$, the transition
to hexagons becomes transcritical and hexagons are stable to stripes
for a $\gamma-$range given by\[
\gamma_{c}-\frac{\sigma^{2}}{4\lambda(b_{0}+2b_{2})}<\gamma<\gamma_{c}+\frac{\sigma^{2}(2b_{0}+b_{2})}{\lambda(b_{2}-b_{0})^{2}}{\equiv\gamma}_{HS}.\]
Note that $\gamma_{HS}>\gamma_{c}$, even if (\ref{eq:HexCond1},\ref{eq:HexCond2})
are not satisfied, since stripes do not exist for $\gamma<\gamma_{c}$.
The instability of hexagons at $\gamma_{HS}$ only arises if $b_{2}>b_{0},$
that is, if $3\psi+14\chi/9+\phi>0.$ With $\sigma\ne0$ the up-down
symmetry of the amplitude equations (\ref{eq:HexAmpEqs}) is broken
and, depending on the sign of $\sigma$, either up- or down-hexagons
are preferred.

Turning to other planforms, Judd and Silber found that rectangular
planforms cannot be stable at or near the degeneracy point \cite{JuSi00}.
Interestingly, however, they find that while super-hexagons cannot
be stable at the degeneracy point, they can arise in a very small
parameter regime in its vicinity if the conditions \begin{eqnarray}
\phi & > & 0,\label{eq:SHexCond1}\\
-\frac{\phi}{21}< & \psi & <\frac{\phi}{3}\label{eq:SHexCond2}\end{eqnarray}
 are met. They then can be bistable with hexagons. We find that in
our system the condition (\ref{eq:SHexCond1}) cannot be met. Rewriting
$\phi$ in terms of the deviation $\sigma$ from the degeneracy condition,\[
\phi=-\frac{2(1+\beta^{2})}{a\beta(\nu-\mu\beta)}\left(2\sqrt{1+\beta^{2}}\eta_{i}^{2}-\frac{(a+\beta)\eta_{i}}{\sqrt{1+a^{2}}}\sigma\right),\]
 shows that - for small $|\sigma|$ - $\phi$ can be made positive
only by making $\eta_{i}$ small as well ($\eta_{i}=\mathcal{O}(\sigma))$.
Even then  $\phi$ can only be slightly positive, $\phi=\mathcal{O}(\sigma),$
requiring that $\psi=\mathcal{O}(\sigma)$ in order to satisfy (\ref{eq:SHexCond2}).
Under these conditions all cubic coefficients in (\ref{eq:cubic},\ref{eq:b1},\ref{eq:b2})
would become of $\mathcal{O}(\sigma)$ and without knowledge of the
next-order coefficients no stability predictions can be made.

Away from the degeneracy, for $\sigma=\mathcal{O}(1)$, the above
arguments suggest that it should be possible to satisfy the stability
conditions (\ref{eq:SHexCond1},\ref{eq:SHexCond2}). For $\sigma=\mathcal{O}(1)$
they are, however, not the correct stability conditions since they
ignore the angle dependence of the cubic coefficients, which is $\mathcal{O}(\sigma)$.
We use (\ref{eq:SHexCond1},\ref{eq:SHexCond2}) therefore only as
a guide to locate parameter regimes in which super-hexagons may be
expected to be stable and then determine the full stability eigenvalues.
One such case is given by the linear parameters used in Figure \ref{fig:NeutralCurve}
with the nonlinear parameters $\alpha=-1$ and $\eta=e^{i\pi/4}$.

Moreover, for $\sigma=\mathcal{O}(1)$ the weakly nonlinear analysis
predicts also stable rectangle patterns satisfying $|b_{1}(\theta)|<b_{0}$
if \[
\alpha>|\beta|\left[1-\frac{1+a^{2}}{6a\sqrt{1+\beta^{2}}}\left(\frac{76}{9}-f(\theta)\right)\chi\right].\]
Within the hexagon sub-space, for $\left|b_{2}/b_{0}\right|>1$ hexagons
become unstable to stripes as $\gamma$ is increased beyond $\gamma_{HS}$.
Satisfied simultaneously, the two conditions $\left|b_{2}/b_{0}\right|>1$
and $|b_{1}(\theta)|<b_{0}$ would yield a situation in which hexagons
are unstable to stripes, which in turn are unstable to rectangle patterns.

Using direct numerical simulations of the forced complex Ginzburg-Landau
equation (\ref{eq:cgle_123forcing}) we have studied to what extent
the predictions of the weakly nonlinear analysis are borne out. In
the degenerate case $\sigma=0$ we find, as predicted, either stripes
or hexagons to be stable depending on the value of $\alpha$ chosen.
A typical hexagonal pattern obtained from random initial condition
is shown in Fig.\ref{fig:up-and-down}. As expected, it exhibits competing
domains of up- and down-hexagons.

\begin{figure}
\includegraphics[width=7cm,keepaspectratio]{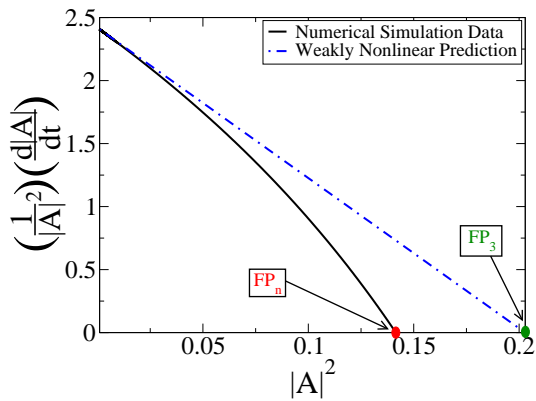}

\caption{Comparing amplitude from numerical simulation with amplitude predicted
by weakly nonlinear analysis. Simulation parameters: $\mu=-1,$$\nu=4$,
$\beta=3$, $\alpha=-1$, $\eta=e^{i\pi/4}$. $FP_{n}$ corresponds
to the fixed point obtained from numerical simulation, $FP_{3}$ to
that obtained from the weakly nonlinear calculation to cubic order.\label{fig:ComparingAmplitude}}
\end{figure}

Away from the degeneracy, $\sigma=\mathcal{O}(1)$, the validity of
the weakly nonlinear analysis can be severely restricted by the fact
that the amplitudes of all stable branches are $\mathcal{O}(1)$,
which formally suggests the significance of higher-order terms in
the expansion. Indeed, in and near the parameter regimes for which
the weakly nonlinear analysis predicts stable super-hexagon patterns
we do not find any indication of their stability. To assess explicitly
the significance of the higher-order terms in the amplitude equations
for $\sigma=\mathcal{O}(1)$ we extract them directly from the numerical
simulation for the case of hexagon patterns. Fig. \ref{fig:ComparingAmplitude}
shows the numerically determined dependence of $|A|^{-2}\, d|A|/dt$
on $|A|$ for $\gamma=\gamma_{c}$. For very small $|A|$ it agrees
well with the result obtained by the third-order weakly nonlinear
theory, which yields the straight dashed line. However, even for $\gamma=\gamma_{c}$
the amplitude $|A|$ saturates only at a value of $|A|\approx0.14$
for which the third-order theory deviates significantly from the full
result. A fit of $|A|^{-2}\, d|A|/dt$ to a higher-order polynomial
shows that the magnitude of the quintic term in the amplitude equation
reaches a value of 20\% of the cubic term. The fixed point obtained
from weakly nonlinear analysis to cubic truncation deviates from the
numercially obtained fixed point by 30\%. This supports out interpretation
that in this regime the cubic amplitude equation does not allow quantitative
predictions.

In summary, we have investigated the regular spatial planforms that
can be stably excited in a sytem undergoing a Hopf bifurcation by
applying a periodic forcing function that resonates with the second
and third harmonic of the Hopf frequency. We have done so within the
weakly nonlinear regime by deriving the appropriate amplitude equations
describing the selection between various planforms. By tuning the
phase of the forcing close to $3\omega_{h}$ one can always reach
the point of degeneracy where no quadratic terms arise in the amplitude
equations, despite the quadratic interaction in the underlying complex
Ginzburg-Landau equation. Hexagons arise then in a supercritical pitchfork
bifurcation. Depending on the parameters of the unforced system we
find that in this regime either the hexagon patterns or stripe patterns
can be stable. In the former case competing domains of up- and down-hexagons
are found in numerical simulations when starting from random initial
conditions. Surprisingly, despite the extensive control afforded by
the two forcing terms, no square, rectangle, or super-hexagon patterns
are stable in the vicinity of this degeneracy, irrespective of the
parameters of the unforced system. Only in the regime in which hexagons
arise in a strongly transcritical bifurcation the weakly nonlinear
theory predicts the possibility of stable rectangles or super-hexagons.
There, however, direct numerical simulations of the complex Ginzburg-Landau
equation indicate no such stability and we show that terms of higher
order in the amplitudes are relevant.

By introducing a further forcing frequency, which is also close to
the 2:1-resonance, the transcritical bifurcation to hexagons can be
avoided. As we show in a separate publication, the corresponding,
more elaborate weakly nonlinear theory then correctly predicts stable
quasi-patterns comprised of four, five, or more modes \cite{CoRi07}.

For larger values of the forcing the spatially periodic standing wave
modes interact with a spatially homogeneous oscillation that is also
excited by the forcing. The interaction between these two types of
modes could lead to interesting patterns, which are, however, beyond
the scope of this paper.

We gratefully acknowledge support by NSF through grant DMS-0309657.

\bibliography{AllPapers,journal}

\end{document}